# Resolving the Bubble Puzzle: Hydrogen Peroxide Formation Precedes Hydroxyl Radicals in Microbubbles and is Governed by Solid–Water Interfaces

Muzzamil Ahmad Eatoo[1,2], Abdul Hamid Emwas[3], Najeh Kharbatia[2], and Himanshu Mishra[1,2,*]

[1]Environmental Science and Engineering (EnSE) Program, Biological and Environmental Science and Engineering (BESE) Division, King Abdullah University of Science and Technology (KAUST), 23955-6900, Thuwal, Kingdom of Saudi Arabia

[2]Water Desalination and Reuse Platform (WDRP), Biological and Environmental Science and Engineering (BESE) Division, King Abdullah University of Science and Technology (KAUST), 23955-6900, Thuwal, Kingdom of Saudi Arabia

[3]Core Labs, King Abdullah University of Science and Technology (KAUST), Thuwal, 23955-6900, Saudi Arabia

*Correspondence: himanshu.mishra@kaust.edu.sa

**Graphical Abstract (TOC)**

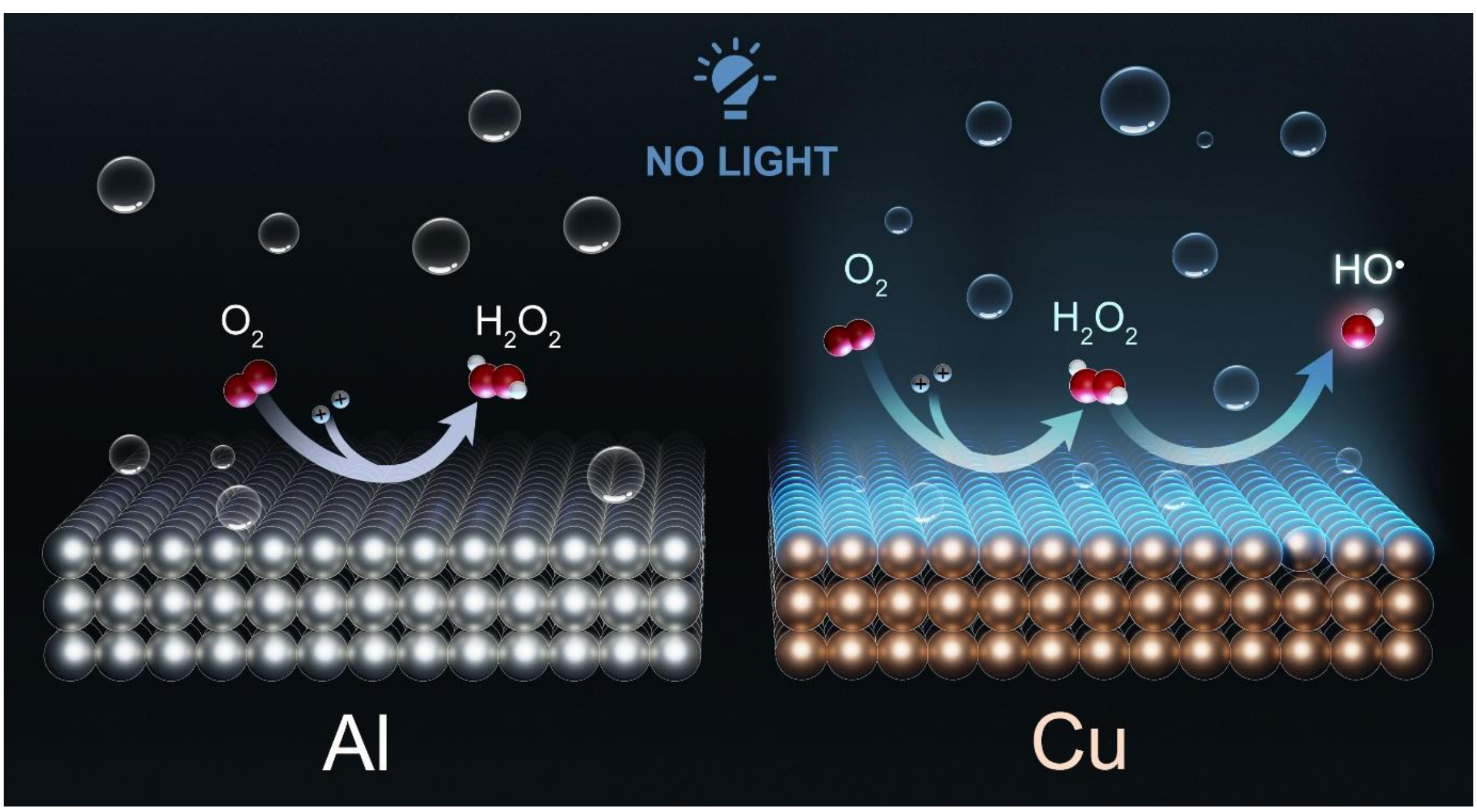

**Abstract**

An alternative explanation is presented for recent reports that attribute sustained chemiluminescence (CL) and electrochemiluminescence (ECL) from electrogenerated microbubbles on steel or copper electrodes in aqueous luminol solutions (over 2–30 V range) to the spontaneous formation of $HO^•$ radicals at the gas–water interface. Our experiments with a broad set of electrodes, viz., stainless steel, copper, aluminium, and platinum, reveal that while microbubbles can be electrogenerated on all electrodes, CL/ECL is exhibited by steel and Cu only and not by Al and Pt. These observations establish that the gas–water interface of microbubbles is not the site for $HO^•$ radical generation (else CL/ECL would be recorded in all cases). Complementary quantification of $H_2O_2$ in these experiments reveals its electrode dependence as follows: Al > Cu > Steel >> Pt (undetectable, limit of detection 50 nM). This establishes that depending on the electrode, $H_2O_2$ forms first, and in some cases, $HO^•$ radicals are observed (i.e., where CL/ECL is seen). Experiments with $^1$H-NMR and electron paramagnetic resonance (EPR) spectroscopy revealed that: (i) $H_2O_2$ formation occurs only when $O_2$ is present in water (from the air or electrogenerated); and (ii) while steel and copper generate $HO^•$ radicals through one-electron reduction of $H_2O_2$, Al does not promote one-electron reduction of $H_2O_2$ to generate $HO^•$ radicals, and Pt preferentially promotes disproportionation of $H_2O_2$ to $H_2O$ and $O_2$. In fact, we demonstrate that $H_2O_2$ and $HO^•$ radicals can be observed at specific metal–water interfaces even without microbubbles, confirming that the solid surface is the reactive site. Therefore, this work affords electrode-based predictions of whether or not electrogenerated microbubbles would yield CL/ECL in luminol solutions and calls into question the notion of spontaneous formation of $HO^•$ radicals at gas–water interfaces.

## Introduction

Known to play a central role in the Earth's environmental chemistry, hydrogen peroxide ($H_2O_2$) is among the most environmentally benign oxidants with water as its degradation product[1 2, 3 4]. Millions of tons of $H_2O_2$ are manufactured to support a wide array of industries, including bleaching of paper and pulp, textile, and hair[5], municipal wastewater treatment and contaminated soil remediation[6], clean-in-place fouled reverse osmosis membranes in desalination plants, "piranha" surface cleaning of semiconductor wafers[5], medical disinfection[7], and epoxidation and hydroxylation in chemical syntheses[8, 9]. Nearly 95% of the $H_2O_2$ production is manufactured via the anthraquinone auto-oxidation process[8] – pioneered some 90 years ago – that is energy-intensive, requires Pd catalysts and organic solvents, demands centralized infrastructure with hazardous transport of concentrated oxidizer, and suffers from the successive loss of process efficiency due to the accumulation of inactive byproducts. Notably, different industrial applications require different concentrations of $H_2O_2$, for example, desalination requires 10–1,000 ppm (or 0.3–30 mM) scale, while bleaching and wastewater treatment require 35–50% by wt. (or 10–15 M). Considering industrial demands and the abovementioned hazmat challenges storing, handling, and transporting $H_2O_2$, it is therefore desirable to realize greener, decentralized on-demand manufacturing, such as those based on electrochemical $O_2$ reduction[10] and photocatalysis[11].

In recent years, a growing body of experimental work has suggested that 1 ppm-scale (29.4 µM) $H_2O_2$ may form spontaneously at the air–water interface of microdroplets generated by pneumatic spraying or condensation[12-17]. Among the air–water interface-specific mechanisms put forth to explain this unusual microdroplet chemistry include: (i) the presence of ultrahigh electric fields that drive the reaction: $HO^- \rightarrow HO^\bullet + e^-$ (Refs.[12-14, 18, 19]), (ii) the spontaneous formation of water radical cations and hydrated electrons: $H_2O \rightarrow H_2O^{\bullet+} + e^-$ (Ref.[20]), and (iii) formation of charged droplets during spraying[21], followed by dehydration of interfacial hydronium and hydroxide ions leading to $HO^\bullet$ and $H^\bullet$ radicals, which form $H_2O_2$ and $H_2$[22, 23].

These claims have been contested, and alternative explanations have been presented[24-28]. It has been demonstrated that the air–water interface has no bearing on the $H_2O_2$ formation; instead, it is the 2-electron reduction of dissolved oxygen in water at solid–water interfaces that generates $H_2O_2$ and oxidizes the solid surface[24-26]. Conversely, if $O_2$ is removed from the water, $H_2O_2$ concentration in sprayed microdroplets (or bulk) becomes undetectable (limit of detection

≥50 nM, i.e., approx. 3 orders of magnitude below the claims). The primary role of the solid is further confirmed by the strict dependence of $H_2O_2$ concentration in microdroplets condensed on solid substrates – varying over two orders of magnitude – depending on their position in the classical Galvanic series[26]. Lastly, the role of ultrasonication – often exploited in generating microdroplets – as an unintended source of $H_2O_2$ formation has also been pinpointed[28-30]. Complementarily, surface-specific surveys of the air–water interface via vibrational sum frequency generation[31] and quantum chemical calculations[30, 32-35] have found no evidence for static or fluctuating electric field build up. Mass spectrometric studies[36-39] and electrochemical studies[30] have called into question the interpretations of prior experimental results. Yet, the debate is not over (See Refs.[18, 40-63]).

A new report[64] suggests that the gas–water interface of microbubbles (i.e., inside-out microdroplets) also spontaneously generate $HO^{\bullet}$ radicals, which combine to form $H_2O_2$ in accordance with mechanism#1 listed above. The study recruited luminol, which is known to yield chemiluminescence (CL) and electrochemiluminescence (ECL) in the presence of $HO^{\bullet}$ radicals but not in the presence of $H_2O_2$ alone. Microbubbles were generated in aqueous luminol solutions (5 mM in a 0.1 M phosphate buffer at pH 9) on steel or copper electrodes by applying a DC potential in the range of 2–30 V. Representative experiments entailed applying a 24 V stimulus to the electrode for 10 seconds and then turning it off for ~20 seconds, while collecting optical images of the CL/ECL. The observed correlations between the applied voltage, microbubble density, and CL/ECL intensity, were taken as evidence for the gas–water interface to be the site for the formation of $H_2O_2$ via the formation of $HO^{\bullet}$ radicals, validating prior claims with microdroplets[12-14]. As well, $H_2O_2$ formation was confirmed by a fluorescent probe for the hydrogen peroxide assay, Amplex Red (10-acetyl-3,7-dihydroxyphenoxazine), in the presence of the horseradish peroxidase enzyme. The conclusion of this study has been corroborated by another group that utilized a microbubble generator in aqueous luminol solutions and applied fluorescence and chemiluminescence microscopy and quantum mechanical/molecular modeling[65]. Together, they suggest mechanism#1 to be at play, while the first study also suggests mechanism#3 *"involving electron transfer from* $OH^{-}$ *to* $H^{+}$ *to form* $H^{\bullet}$ *and* $HO^{\bullet}$ *is also important"*[64]. While these mechanistic pathways are disparate, they agree on the hypothesis that $HO^{\bullet}$ radicals are formed first at the gas–water interface and then they recombine to form $H_2O_2$.

Prior studies are also silent on the role of the underlying electrode beyond it being the site for generating microbubbles, which has been pointed out earlier[25, 26]. We wondered if the observed chemistry arises purely from the gas–water interface of microbubbles or from electrode-dependent interfacial processes. With this backdrop, here we re-examine the origins of $HO^{\bullet}$ radicals and $H_2O_2$ – including their formation sequence – in electrogenerated microbubbles on a wide range of electrodes (stainless steel, copper, aluminium, and platinum) by recording CL/ECL events and quantifying $H_2O_2$ and $HO^{\bullet}$ concentrations via $^{1}$H-NMR and EPR spectroscopy, respectively. The specific questions we address are:

1. Do electrogenerated microbubbles in luminol solutions **always** produce CL/ECL?
    a. If yes, it would confirm that gas–water interfaces generate $HO^{\bullet}$ radicals, which then form $H_2O_2$ by recombination.
    b. If not, then the question would be – Did $HO^{\bullet}$ radicals form first, leading to $H_2O_2$ or vice versa? That is, it would question their temporal sequence of formation.
2. Does the reduction of (electrogenerated) $O_2$ in the microbubbles contribute to $H_2O_2$ formation at the water–electrode interface?
    a. If yes, this mechanism would lead to $H_2O_2$ formation but without yielding CL/ECL.
    b. What would be the relative fraction of this pathway compared to $HO^{\bullet}$ radicals forming $H_2O_2$ in water?
3. Could $H_2O_2$ react at the water–metal interface to form $HO^{\bullet}$ radicals?
    a. If yes, how would this reaction contribute/interfere with the CL/ECL noted in question#1 above?

Our experimental results reveal that CL/ECL events depend on the nature of the electrode, i.e., CL/ECL events do not always accompany microbubbles. We demonstrate that $H_2O_2$ forms first at the electrode–water interface via $O_2$ reduction, then $H_2O_2$ can decompose to form $HO^{\bullet}$ radicals or $HO^{-}$ or $H_2O$ and $O_2$ depending on the nature of the metal surface. An alternative explanation – consistent with our prior contribution – emerges: the solid–water interface governs the chemistry, not the gas–water interface. This explains why the CL/ECL events are dependent on the nature of the electrode.

**Results and Discussion**

Following the experimental setup and protocols reported by Zare & co-workers[64], we investigated the formation of $HO^{\bullet}$ radicals and $H_2O_2$ at the gas–water interface of electrogenerated microbubbles. While prior studies primarily focused on copper and steel electrodes, we systematically expanded the range of electrode materials to include aluminium (AA5083), stainless steel (SS304), copper, and platinum surfaces (See the Methods section for the preparation of surfaces, especially removing the native oxide layer of the metal surfaces prior to the experiment). Similar to the previous study by Zare and coworkers[64], the electrochemical setup consisted of a split closed bipolar electrode setup (Figure 1). One of the chambers was connected to the positive DC terminal, while the other chamber was connected to the negative DC terminal. Each chamber contained two electrodes: one directly connected to the external voltage source and the other electrically linked to the corresponding electrode in the opposite chamber via a connecting wire (See Figure 1). This configuration separated redox processes across distinct electrode surfaces.

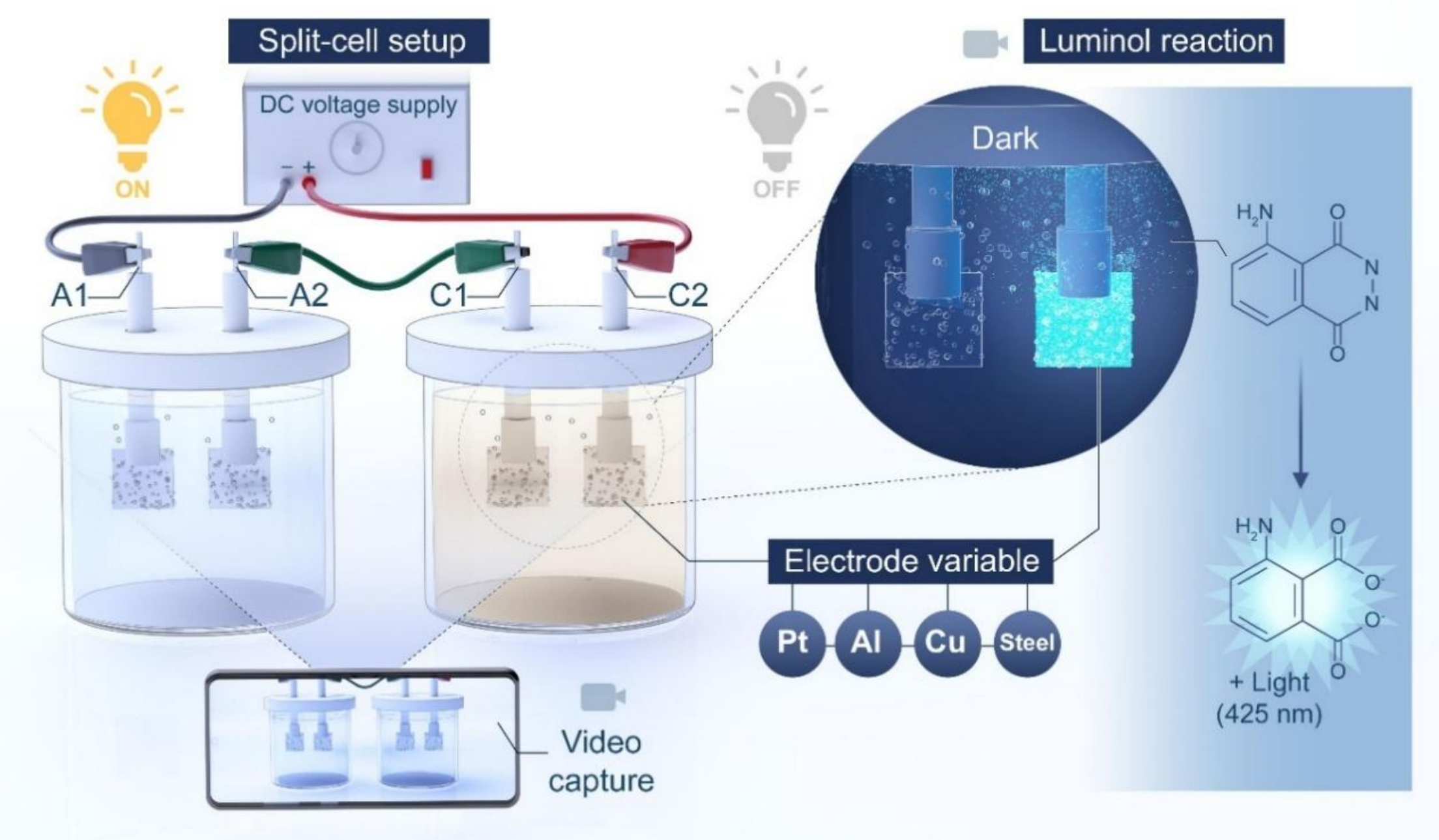

Fig.1: Schematic diagram of the experimental setup of a split closed bipolar electrode setup used to produce bubbles using DC voltage supply. It consists of two chambers; the right chamber contains electrodes C1 & C2 in 5 mM aqueous luminol solution maintained at pH 9 to compare the observations with the prior study[64] and luminol is soluble only in alkaline aqueous solutions. The left chamber contains two electrodes (A1 & A2) in alkaline (pH 9)

solution. The bubbles were generated on electrode surfaces by applying DC voltage and chemiluminescence was recorded in the dark using an optical camera.

Microbubbles were electrochemically generated on aluminium (Al), copper (Cu), platinum (Pt), and stainless-steel electrodes in 5 mM aqueous alkaline luminol solution by applying a 20 V potential for 10 s. Bubble formation was documented under ambient laboratory lighting using a smartphone camera, and chemiluminescence images were acquired in complete darkness 20 s after the applied potential was discontinued. Experimental results revealed that chemiluminescence exhibited a stark dependence on the nature of the electrode material. While we reproduced the observations of Zare & co-workers[64], i.e., CL/ECL accompanied with electrogenerated microbubbles on copper and steel electrodes, this finding was not generalizable. No chemiluminescence was observed when bubbles were generated on aluminium or platinum electrodes, even though the electrogenerated bubbles were clearly observed. (Fig 2 & Supplementary Movies 1–3).

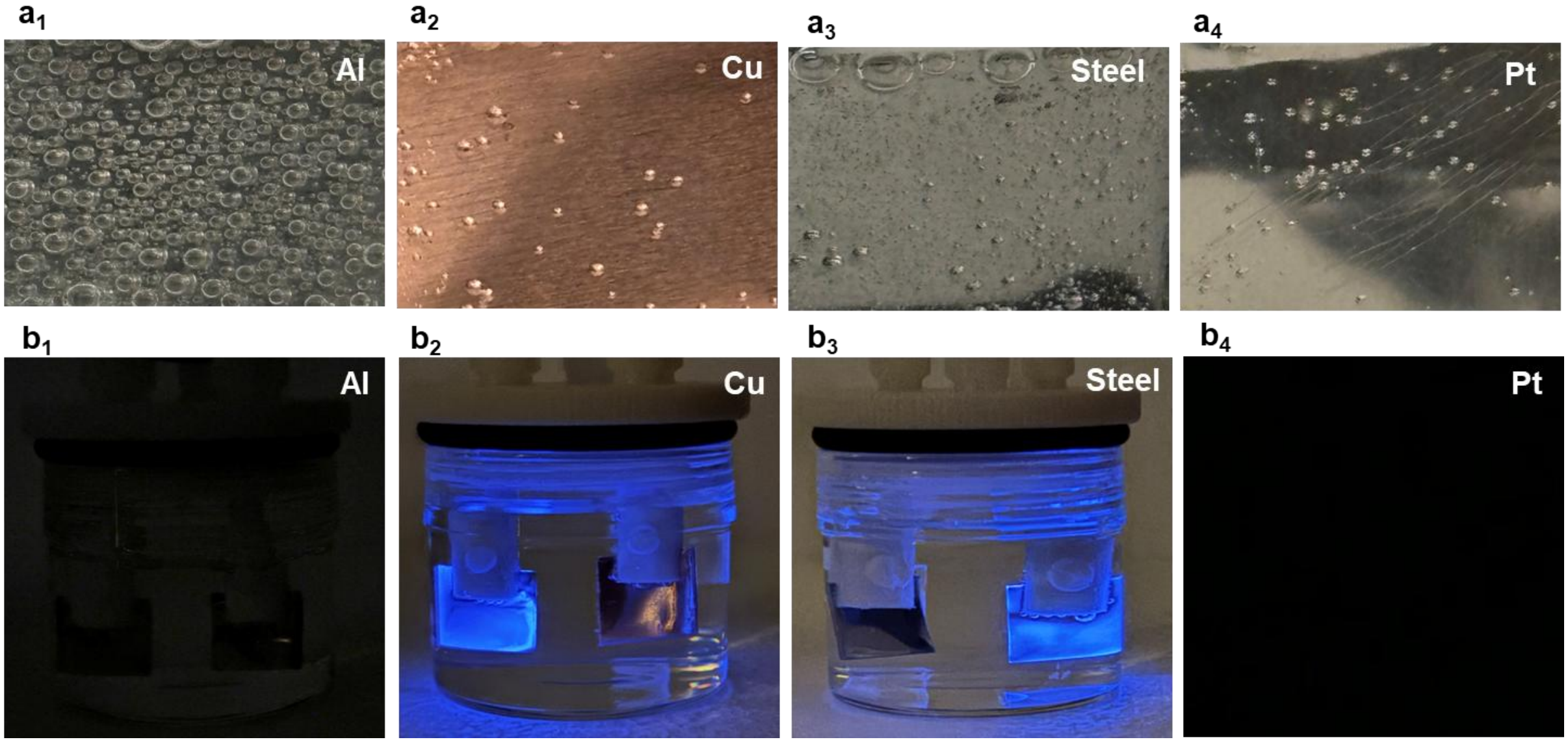

Fig. 2 Electrogenerated microbubbles and CL/ECL events do not always accompany each other. Optical images under light and dark conditions: ($a_1$) Aluminium, ($a_2$) Copper, ($a_3$) Steel, and ($a_4$) Platinum. Corresponding chemiluminescent observed in dark ($b_1$) Aluminium, ($b_2$) Copper, ($b_3$) Steel, and ($b_4$) Platinum.

We repeated these experiments in both two-electrode and three-electrode configurations, yielding consistent results across all setups, thereby confirming that the observed behaviour is independent of cell geometry or measurement configuration. These results demonstrate that CL/ECL is governed by the electrode material and not by the gas–water interface of microbubbles. This result contradicts the claim that $HO^•$ radicals form at gas–water interfaces of microbubbles. If the gas–water interface was the universal source of hydroxyl radicals, all electrodes producing comparable populations of microbubbles should exhibit comparable chemiluminescence. Instead, only Cu and stainless steel emit light, whereas Al and Pt do not, despite readily producing microbubbles. Therefore, the gas–water interface cannot be the universal origin of hydroxyl radicals. **This also raises a critical mechanistic question: why CL/ECL (i.e., $HO^•$ radical formation) is observed exclusively with steel (Fe-based) and copper electrodes, but not with aluminium or platinum, despite comparable bubble generation?** We answer this question after reporting on the observed $H_2O_2$ concentrations in these experiments, as that will aid the understanding.

$H_2O_2$ generation was directly quantified using $^1H$ NMR spectroscopy following a method reported by Bax and coworkers[66]. In this experiment, microbubbles were produced by applying a 20 V potential for 10 s and were subsequently allowed to persist on the electrode surfaces for 1 h before analysis. The signals corresponding to $H_2O_2$ formed at different metal electrodes are presented in Fig. 3. These NMR measurements were performed at 275 K on a 950 MHz Bruker Avance Neo spectrometer equipped with a 5 mm z-gradient TCI cryogenic probe, employing a 6 ms Gaussian 90° pulse for selective excitation of the $H_2O_2$ proton resonance (see Methods section).

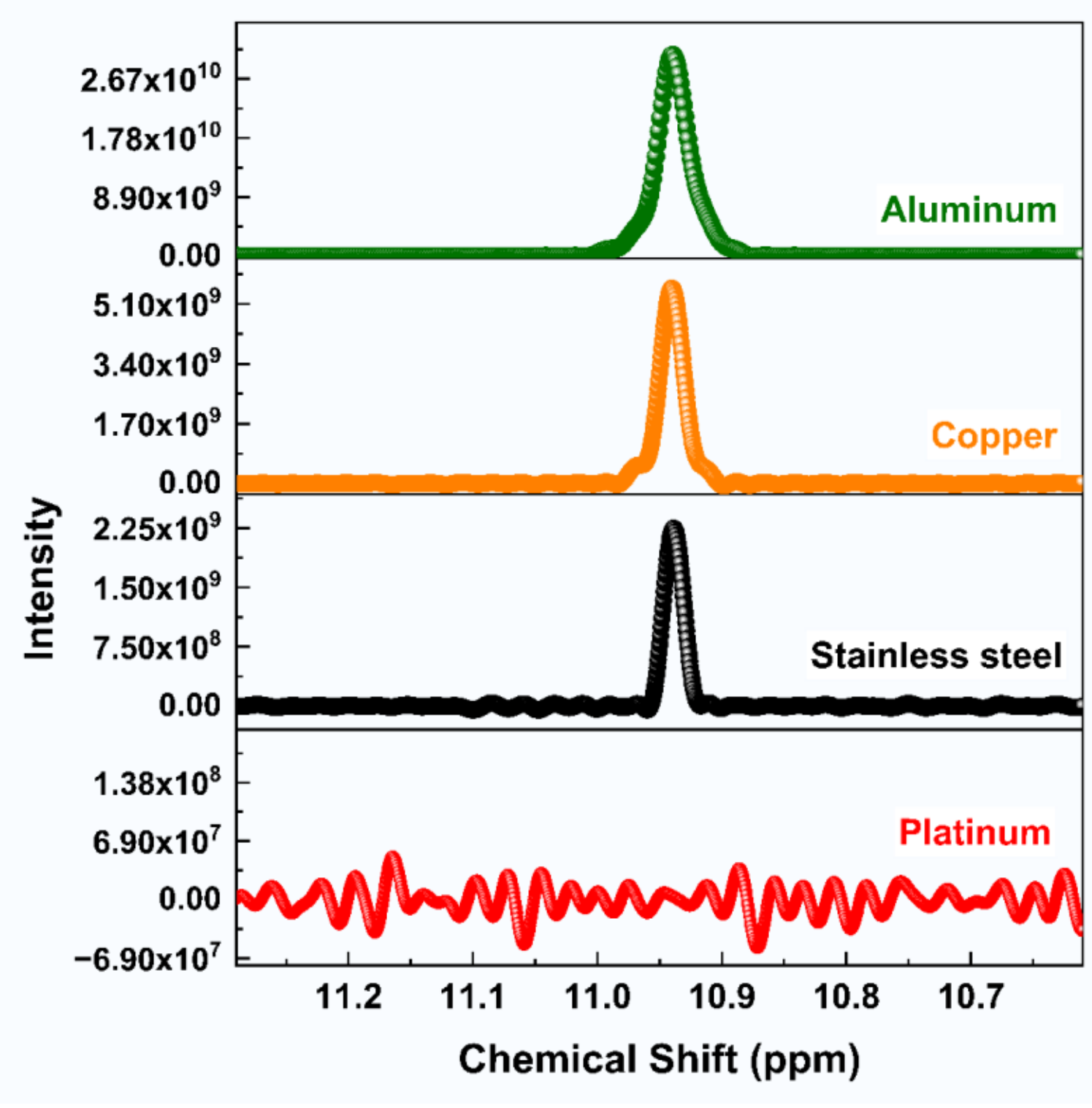

Fig. 3 Electrogenerated microbubbles and concomitant $H_2O_2$ formation. Microbubbles are generated by applying voltage, and then after 1 h, samples are analyzed by $^1$H NMR for $H_2O_2$ concentrations. These concentrations vary with the nature of electrode (Al, Cu, Pt, and stainless steel) as: Al > Cu > stainless-steel >> Pt (undetectable).

It turns out that the aluminium electrode, despite exhibiting no detectable chemiluminescence during bubble formation, produced substantially higher concentrations of $H_2O_2$ (9.5 ± 0.4 μM) than the Cu (1.6 ± 0.2 μM) and stainless-steel (0.7 ± 0.2 μM) electrodes, both of which displayed strong chemiluminescence. In contrast, no $H_2O_2$ was detected for the Pt electrode, with concentrations remaining below the NMR detection limit of 50 nM. The absence of $H_2O_2$ (Figure 3) and $HO^\bullet$ radicals (chemiluminescence/Figure 2) in the case of bubbles formed on platinum electrode reveals that gas–water interface of microbubbles does not produce these reactive oxygen species.

The findings presented in Figs. 2 and 3 indicate that $H_2O_2$ and $HO^\bullet$ radicals are not generated at the gas–water interface of microbubbles. Rather, they depend on the nature of the electrode – and therefore signal electrochemical processes at play. To rigorously test this, we simplified the system by eliminating bubble formation altogether and examined $H_2O_2$ and $HO^\bullet$ generation at the solid–water interfaces of Al, Cu, Pt, and stainless steel. Freshly polished metal specimens with the same surface area (2.4 $cm^2$) were immersed in equal volume of alkaline (pH 9) water (2 mL) for one hour, and interfacial $H_2O_2$ production was quantified by high-sensitivity $^1$H NMR spectroscopy. Even in the complete absence of microbubbles, spontaneous $H_2O_2$ formation was readily detected at the metal–water interfaces. Notably, the Al–water interface produced substantially higher concentrations of $H_2O_2$ than the Cu–water and stainless-steel–water interfaces in oxygen-containing water (Fig. 4a; see Methods for analytical details), whereas no detectable $H_2O_2$ was observed for Pt. Quantitatively, the $H_2O_2$ production rates in these experiments followed the trend: Al (28 μM) ≫ Cu (2.4 μM) > stainless steel (1.5 μM). These findings establish the metal–water interface as the site of $H_2O_2$ generation and reveal that both the rate and extent of $H_2O_2$ production are critically dependent on the identity of the metal.

Notably, these findings (Figs. 2 & 3) also demonstrate that the presence of $H_2O_2$ alone is insufficient to produce chemiluminescence in a luminol solution. Although the Al electrode generated the highest concentration of $H_2O_2$, no chemiluminescence was observed during bubble formation (Fig. 2). This conclusion was further validated by directly adding $H_2O_2$ to a luminol solution under identical experimental conditions (5 mM luminol, pH 9), which likewise produced no detectable chemiluminescence. These results indicate that $H_2O_2$ itself is not the emissive species and that additional reactions at specific electrode surfaces are required to generate the reactive intermediates responsible for luminol chemiluminescence.

**Next, we sought to answer the question: why CL/ECL (i.e., HO$^•$ radical formation) is observed exclusively with steel (Fe-based) and copper electrodes but not with aluminium or platinum, despite Al producing significantly higher $H_2O_2$**. The spontaneous formation of HO$^•$ radicals at these metal–water interfaces was directly investigated by electron paramagnetic resonance (EPR) spectroscopy. Metal specimens of Al, Cu, Pt, and stainless steel were immersed in aqueous solutions under identical conditions (constant surface area, solution volume, pH (9), and composition (10 mM DMPO solution)), and the generation of HO$^•$ radicals was investigated by EPR spectroscopy using 5,5-dimethyl-1-pyrroline *N*-oxide (DMPO) as a spin-trapping agent.

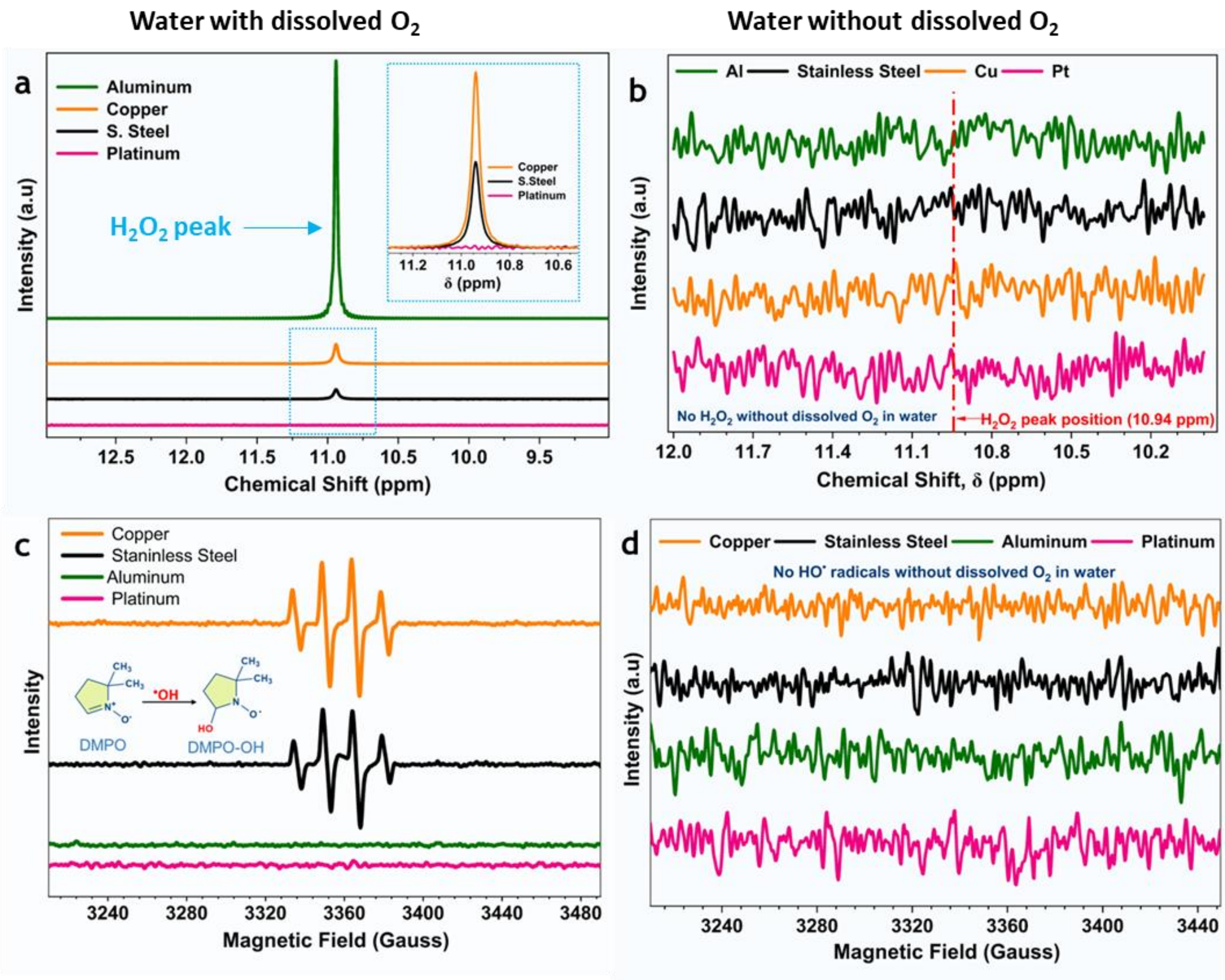

Fig. 4 Spontaneous formation of $H_2O_2$ and HO$^•$ at the solid–water interface. (a) $^1$H-NMR spectra showing the comparison of $H_2O_2$ formed at different metal–water interfaces in presence

of dissolved oxygen. (b) $^{1}$H-NMR spectra showing no detectable (detection limit 50 nM) $H_2O_2$ forms at the metal–water interfaces in absence of dissolved oxygen. (c) EPR spectroscopy results showing spontaneous formation of DMPO-OH or $HO^{\bullet}$ radical at different metal–water interfaces in presence of dissolved oxygen. (d) EPR spectroscopy results showing no detectable DMPO-OH or $HO^{\bullet}$ formation at the metal–water interfaces without $O_2$.

The EPR spectroscopy measurements (Fig. 4c) revealed that, among the examined metals, only Cu and stainless steel (Fe) spontaneously generate $HO^{\bullet}$ radicals. This observation provides a compelling explanation for the chemiluminescence observed during microbubble formation in luminol solution (Fig. 2), where light emission occurs exclusively with Cu and Fe-containing electrodes. In contrast, although Al produced the highest concentration of $H_2O_2$ (Fig. 4a), no detectable $HO^{\bullet}$ radicals were observed, demonstrating that $H_2O_2$ generation alone is insufficient to produce hydroxyl radicals or induce luminol chemiluminescence. To establish the origin of these reactive oxygen species, the experiments were repeated in oxygen-free water inside an $N_2$-filled glovebox. Under these conditions, neither $H_2O_2$ nor $HO^{\bullet}$ radicals were detected (Figs. 4b,d) for any of the investigated metals, confirming that dissolved $O_2$ is the precursor for both $H_2O_2$ and $HO^{\bullet}$ radicals. Collectively, these results demonstrate that dissolved $O_2$ first reduces at the metal–water interface to form $H_2O_2$. The subsequent fate of $H_2O_2$ is dictated by the nature of the metal surface: Cu and Fe-containing surfaces enable its conversion into $HO^{\bullet}$ radicals through Fenton-like one-electron reduction, whereas Al promotes $H_2O_2$ formation but does not promote one-electron reduction of $H_2O_2$ to form $HO^{\bullet}$ radicals. This mechanistic distinction explains why chemiluminescence is observed only for Cu and Fe-containing electrodes despite the substantially higher $H_2O_2$ production on Al.

To establish the relationship between $H_2O_2$ generation from oxygen reduction reaction and the underlying interfacial oxidation processes, potentiodynamic polarization measurements were employed using a conventional three-electrode electrochemical configuration under identical solution conditions. Potentiodynamic polarization provides direct information on the oxidation (anodic) behavior and corrosion kinetics of a metal surface by measuring the current response as the electrode potential is systematically varied. Consequently, it enables quantitative evaluation of the tendency of different metals to undergo interfacial oxidation. In this experiment, the metals under examination (Al, Cu, and stainless steel) served as the working electrode, while a platinum electrode and a saturated calomel electrode were used as the counter and reference electrodes, respectively. The working electrode potential was swept from -0.5 to +1.0 volts with scan rate of 0.5 mV/s and the resulting polarization curves (Fig. S1) were used to determine the anodic oxidation behavior of each metal. The corresponding polarization curves are presented in Fig. S1 Notably, the corrosion current densities or rates ($i_{corr}$) exhibit the same trend as the $H_2O_2$ production rates: Al (320 μA/cm$^2$) ≫ Cu (25 μA/cm$^2$) > stainless steel (16 μA/cm$^2$). The ratios between $H_2O_2$ production and corrosion currents match – Al : Cu : Steel :: 19 : 1.6 : 1 – providing compelling evidence that $H_2O_2$ formation is governed by electrochemical oxidation of the solid surface coupled to oxygen reduction at the solid–water interface, rather than processes occurring at the gas–water interface of microbubbles.

**The $HO^\bullet$ radicals detected in Fig. 4c originate from the one-electron reduction of $H_2O_2$ that was spontaneously formed from $O_2$ reduction at the solid–water interface. However, the spontaneous $H_2O_2$ concentrations at different metal–water interfaces (Fig. 4a) are not identical because $H_2O_2$ formation is governed by the rate of dissolved $O_2$ reduction, which depends on the oxidation characteristics of the metal surface.** As established in our previous studies[25, 26] and this study as well, the tendency of different metals to generate $H_2O_2$ follows the classical Galvanic series and requires dissolved $O_2$, a finding further confirmed by the absence of both $H_2O_2$ and $HO^\bullet$ formation under oxygen-free conditions (Fig. 4). Consequently, directly comparing $HO^\bullet$ formation under spontaneous conditions would conflate differences in $H_2O_2$ production with differences in the intrinsic reactivity of each metal to decompose $H_2O_2$. To isolate the latter process, identical concentrations of $H_2O_2$ were added externally while dissolved $O_2$ was eliminated to suppress any further interfacial $H_2O_2$ generation. Metal pellets (Al, stainless steel, Cu, Au, and Pt) were then immersed in 10 mM $H_2O_2$ solution in sealed vials. $HO^\bullet$ radical formation was monitored by electron paramagnetic resonance (EPR) spectroscopy using DMPO as the spin trap. This experimental design enables a direct comparison of the intrinsic ability of different metal surfaces to catalyze the conversion of $H_2O_2$ into $HO^\bullet$ radicals.

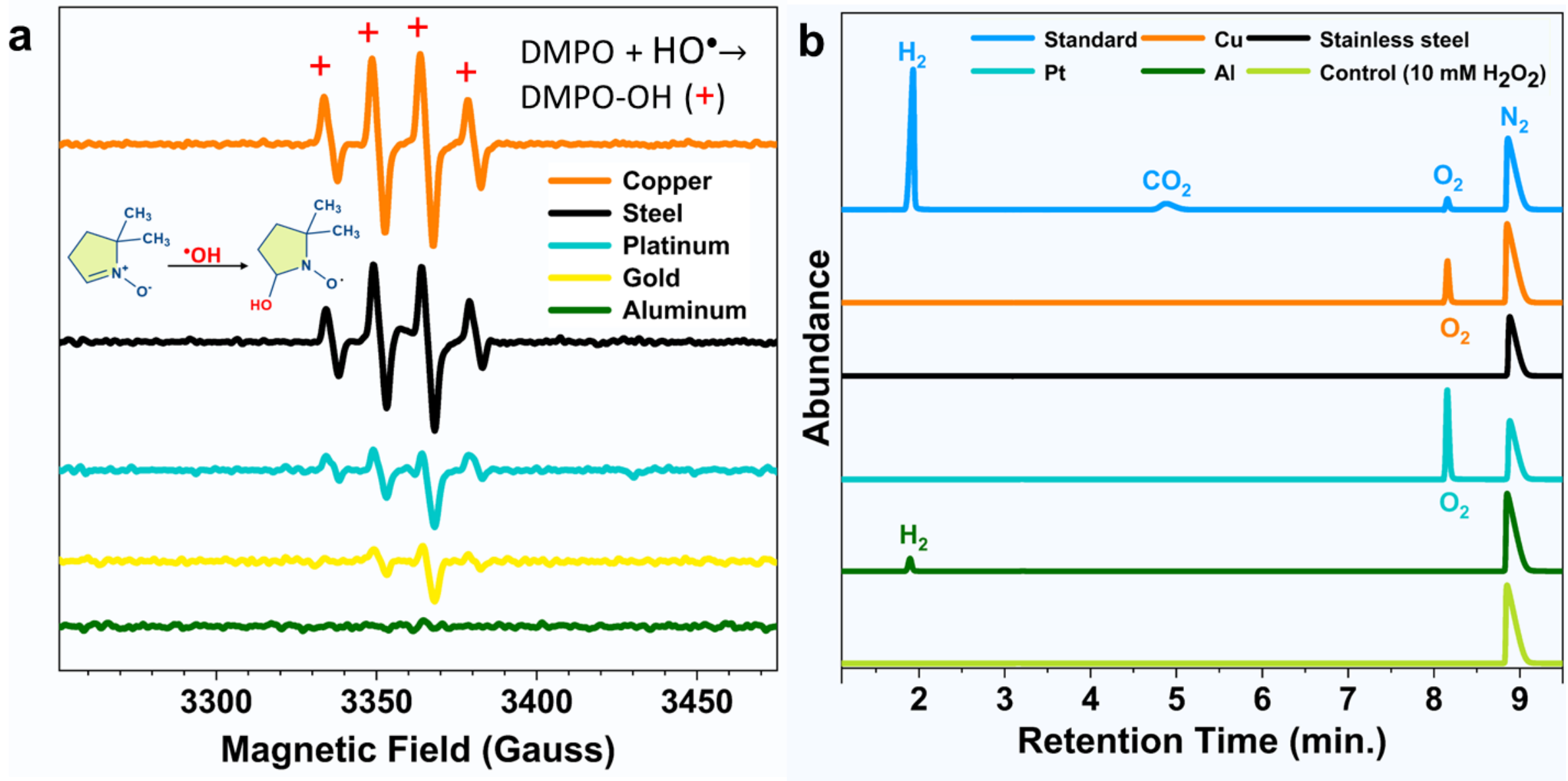

**Fig. 5** Comparative decomposition of aqueous $H_2O_2$ (10 mM) at metal surfaces. **a** Formation of $HO^{\bullet}$ radicals at Cu, steel, Pt, Au, and Al surfaces reveals the following trend: Cu > stainless steel >> Pt > Au >> Al (no detectable DMPO-OH). **b** Evolution of $O_2$ gas in the vial headspace containing metal pellets in a 10mM aqueous $H_2O_2$ solution.

The results (Fig. 5a) revealed that, among the materials investigated, Cu generated the highest $HO^{\bullet}$ radicals from $H_2O_2$, while Al did not produce any detectable amount. The comparative tendency of these metals to generate $HO^{\bullet}$ radicals from $H_2O_2$ follows the order Cu > stainless steel >> Pt > Au >> Al (no detectable DMPO-OH). Interestingly, it also shows that if $H_2O_2$ is present in water, even noble materials like platinum and gold can form $HO^{\bullet}$ radicals. Additionally, these experiments also establish that $H_2O_2$ is indeed an intermediate capable of generating $HO^{\bullet}$ radicals at some solid–water interfaces. Further, the evolution of gaseous products during $H_2O_2$ decomposition on different metals was examined by gas chromatography (GC). The metal samples were immersed in 10 mM $H_2O_2$ solution for 24 hours and the gas-phase products were analyzed using an Agilent 7890A gas chromatograph equipped with thermal conductivity detector (TCD), micro electron capture detector (μECD), and sulfur chemiluminescence detector (SCD) detectors. Gases ($N_2$, $O_2$, $CO_2$, $H_2$, and $CH_4$) were separated and quantified using the TCD channel with argon as the carrier gas, employing a ten-port sampling valve for water back-flushing and a six-port valve for column switching. Headspace gas was collected from sealed sample bottles through a silicone septum using a 2.5 mL gas-tight syringe and injected directly into the GC. The GC results (Fig 5b) revealed that platinum produces significantly higher $O_2$ gas from $H_2O_2$ whereas stainless steel and aluminium do not show $O_2$ evolution. Interestingly, at higher $H_2O_2$ concentration (millimolar and above range), Cu also shows some $O_2$ gas evolution. Notably, GC results showed that aluminium produces $H_2$ gas.

## Discussion

The combined use of EPR, $^1$H-NMR, luminol chemiluminescence, and potentiodynamic polarization measurements enabled us to distinguish between two competing mechanistic proposals: (i) formation of $HO^•$ radicals at the gas–water interface followed by recombination to yield $H_2O_2$, or (ii) formation of $H_2O_2$ via reduction of dissolved $O_2$ at the solid–water interface, followed by $HO^•$ radical formation. Here, we assimilate the results to advance a new interpretation for the CL and ECL observed in microbubbles (electrogenerated or not) in aqueous luminol solutions. **Our results establish that as long as dissolved $O_2$ is present in aqueous solutions – sourced from electrogenerated $O_2$ or simply from the ambient air – $H_2O_2$ will form on common material surfaces such as Al, Cu, and stainless steel in those solutions due to the redox reaction of dissolved $O_2$.** This study revealed that electrogenerated microbubbles (containing $O_2$ gas) do not guarantee CL/ECL which disproves the prior suggestions[64, 65] that $H_2O_2$ is formed at the air–water interface due to the $HO^•$ radical recombination (Fig. 2). These findings directly rationalize the electrode-dependent chemiluminescence observed in luminol-containing solutions: strong emission is detected for Cu and steel electrodes, while no emission is observed for Al and Pt electrodes, despite comparable microbubble formation. Luminol chemiluminescence arises from the formation of $HO^•$ from Fenton-like one electron reduction of $H_2O_2$ at the metal surface or the solid–water interface, i.e., there is no $HO^•$ radical formation on microbubbles. Therefore, it is clear that $H_2O_2$ is first produced via oxygen reduction at the solid–water interface, and subsequently converted into $HO^•$ at specific solid surfaces; notably, the resulting metal cations may also react with $H_2O_2$ to generate the hydroxyl radicals. In other words, $HO^•$ radicals are formed from one-electron reduction (Fenton-like) of $H_2O_2$ and are not generated intrinsically at the gas–water interface of microbubbles or microdroplets. The dominant pathway for $H_2O_2$ decomposition at different types of materials is summarized in Figure 6.

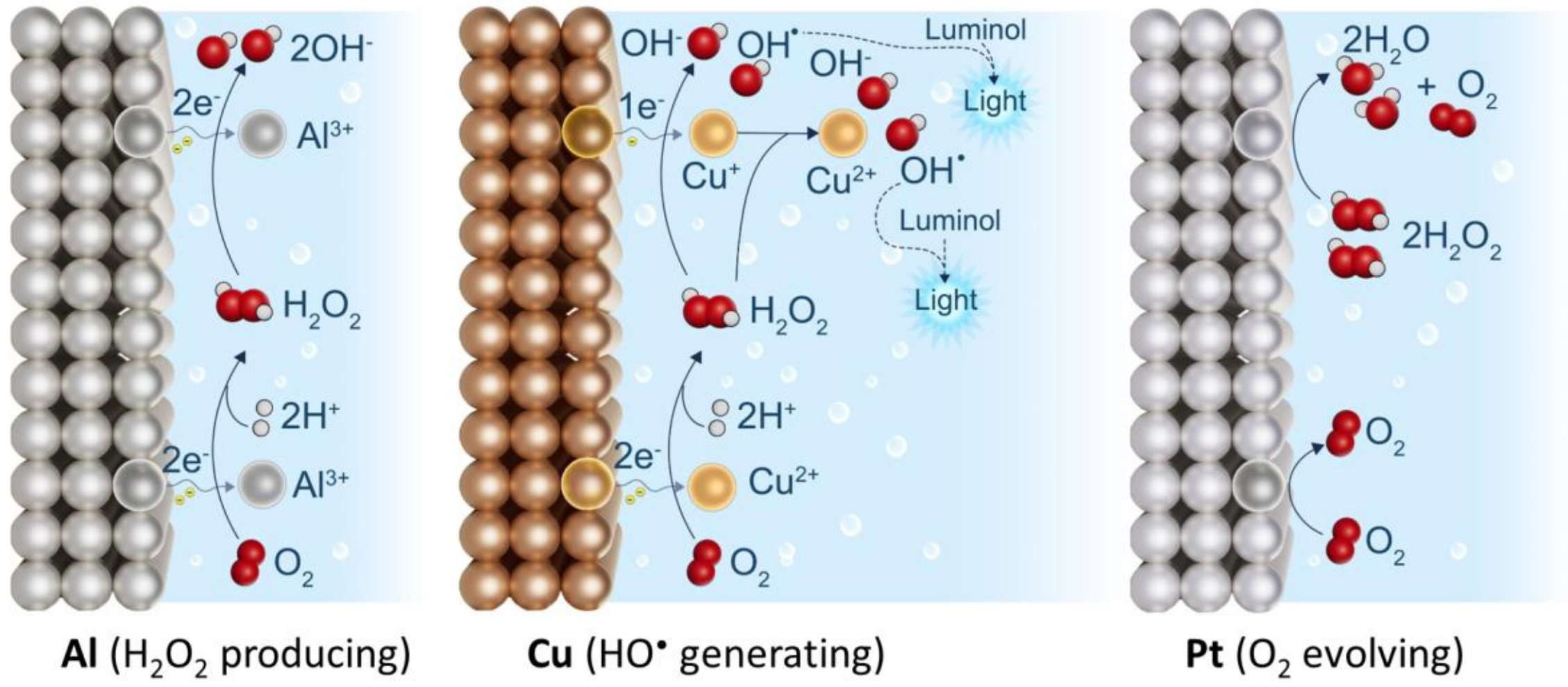

Figure 6. Material–dependent pathway for generating / decomposing $H_2O_2$ at the solid–water interfaces. Illustration shows (a) Al mainly reduces $O_2$ to $H_2O_2$ spontaneously and does not generate any $HO^\bullet$ radicals. (b) Materials like Cu and steel (Fe-based) not only reduce $O_2$ to form $H_2O_2$ but also reduce $H_2O_2$ through Fenton-like one-electron reduction to generate $HO^\bullet$ radicals spontaneously. (c) Materials like Pt does not reduce $O_2$ spontaneously but it enables $O_2$ evolution from $H_2O_2$ through its disproportionation to $H_2O$ and $O_2$ spontaneously.

Thus, each material selectively governs the fate of dissolved oxygen and $H_2O_2$ (Fig. 6), rather than generating $HO^\bullet$ radicals at the gas–water interface through a universal mechanism. For instance, aluminium efficiently drives the 2-electron reduction of dissolved $O_2$ to $H_2O_2$ (Eq. 1) and then promotes two-electron reduction of $H_2O_2$ to form two hydroxide ions, Cu and Fe-containing surfaces reduce dissolved $O_2$ through 2-electron reduction to $H_2O_2$ and subsequently promote 1-electron reduction of $H_2O_2$ into $HO^\bullet$ radicals (Eq. 2), whereas Pt preferentially promotes disproportionation of $H_2O_2$ to $O_2$ and $H_2O$ (Eq. 3). Notably, Pt being a noble material does not produce $H_2O_2$ through the 2-electron reduction of $O_2$, but if $H_2O_2$ is added externally it majorly promote its disproportionation (Eq. 3) and it can also lead to some $HO^\bullet$ radicals formation (Fig. 5a), Thus, the chemical identity of the solid surface determines not only the amount of $H_2O_2$ and $HO^\bullet$ radicals generated but also the specific pathway that dominates at the interface.

**Aluminium** ($H_2O_2$ producing surface)**:**

$$\mathbf{O_2} + 2H^+ + 2e^- \rightarrow \mathbf{H_2O_2} \qquad \text{Eq. (1)}$$

**Copper/stainless steel/Fe** ($HO^\bullet$ forming surface)**:**

$$\mathbf{O_2} + 2H^+ + 2e^- \rightarrow \mathbf{H_2O_2} + e^- \rightarrow \mathbf{HO^\bullet} + HO^- \quad \text{Eq. (2)}$$

**Platinum** ($O_2$ evolving surface)**:** $\mathbf{2H_2O_2 \rightarrow O_2 + 2H_2O}$ Eq. (3)

**Conclusion:**

Contrary to what has been suggested in Refs.[64, 65], we demonstrate that the gas–water interface of microbubbles (electrogenerated or not) neither generates $HO^\bullet$ radicals nor guarantees chemiluminescence (CL). Instead, CL/ECL depends on the chemical composition of the electrode material underlying the bubbles. Our laboratory experiments revealed that as dissolved $O_2$ (or electrogenerated $O_2$) gets reduced at the metal surface, $H_2O_2$ is formed and the metal is oxidized. Next, the fate of $H_2O_2$ depends on the electrode itself, e.g., Cu and steel are efficient at decomposing $H_2O_2$ to form $HO^\bullet$ radicals via Fenton chemistry, while aluminium is inert towards $H_2O_2$ and Pt preferentially decomposes $H_2O_2$ through $O_2$ evolution. This insight explains the disparate CL/ECL responses of electrogenerated microbubbles on different electrodes in luminol solutions (Fig. 2). These findings are consistent with our prior reports on $H_2O_2$ formation in sprayed and condensed water microdroplets, establishing $O_2$ reduction at the solid–water interface as the common origin of $H_2O_2$ (Refs.[24-26]). More broadly, the electrode-dependent outlook summarized in Eqs. 1–3 enables prediction of whether a given electrode will produce CL/ECL in luminol solution, and cautions that any study reporting spontaneous $H_2O_2$ and $HO^\bullet$ radicals formation at aqueous interfaces must account for solid surfaces present in the experimental apparatus.

## Methods

### Chemicals & Materials

Luminol (5-amino-2,3-dihydrophthalazine-1,4-dione) was purchased from Tokyo Chemical Industry (CAS RN: 521-31-3), DMPO (5,5-Dimethyl-1-pyrroline N-oxide, CAS: 7789-20-0) was purchased from Sigma-Aldrich, standard $H_2O_2$ (30%; Cat. 270733), Phosphate buffer, Sodium hydroxide (CAS: 1310-73-2), and HPLC-grade water (Cat. 2594649). Amplex Red Hydrogen Peroxide/Peroxidase Assay Kit (A22188). The DI water was also obtained from a Milli-Q Advantage 10 setup (with a resistivity of 18.2 MΩ cm). Metals and alloys specimens like aluminium (5083), copper, stainless steel (SS304), gold, and platinum.

### Bubble Generation

A split, closed bipolar electrode configuration (Figure 1) was used to investigate $H_2O_2$ and $HO^{\bullet}$ radicals formation during bubble generation. The setup comprised two physically separated compartments connected via a bipolar electrode, enabling independent anodic and cathodic reactions while minimizing cross-interference. The setup comprised four electrodes, with two electrodes placed in each compartment. Steel electrodes were used in the anodic chamber in all experiments, while the cathodic electrode material was systematically varied (copper, aluminium, platinum, or steel). The electrodes in the two compartments were electrically connected to function as a bipolar pair, while an additional pair of electrodes connected to a DC power supply served as the driving electrodes. The anodic compartment contained 0.1 M KOH, while the cathodic compartment contained 5 mM luminol (pH 9). All experiments were performed using alkaline (pH 9) aqueous solutions because luminol is soluble only in alkaline aqueous solution. Unless otherwise stated, experiments were conducted at ambient conditions. The exposed geometric surface area of each electrode was maintained at ~2 $cm^2$, and the applied potential was controlled using a DC power supply (Wanptek KPS305D). The compartment volumes (e.g., 30 mL) and electrode separation (e.g., 2 cm) were kept constant across all experiments. Before each experiment, the metal specimen surfaces were polished using silicon carbide (SiC) or emery papers starting from 100 to 800 grit size followed by cleaning using high-pressure dry nitrogen gas.

**Potentiodynamic polarization tests**

The corrosion behavior of aluminium (Al), copper (Cu), and stainless steel (SS304) specimens in aqueous media (pH 9) was evaluated using potentiodynamic polarization measurements. Prior to each experiment, the metal surfaces were sequentially polished with silicon carbide (SiC) abrasive papers up to 1200 grit to ensure a reproducible finish, followed by thorough cleaning using a stream of high-purity nitrogen ($N_2$) gas to remove residual particulates. All measurements were performed on specimens with an exposed geometric surface area of 1 $cm^2$ using a BioLogic VMP3 potentiostat. A conventional three-electrode configuration was employed, consisting of the metal specimen as the working electrode, a platinum sheet (1 $cm^2$) as the counter electrode, and a saturated calomel electrode (SCE) as the reference electrode. Polarization scans were carried out at a sweep rate of 0.5 mV $s^{-1}$. The open circuit potential (OCP) was recorded for a brief period (5 s) prior to polarization to limit surface alteration during equilibration. The polarization experiments were then initiated at −500 mV relative to the measured OCP.

**NMR spectroscopy for $H_2O_2$(aq) quantification**

All NMR experiments were performed on a 950 MHz Bruker Avance Neo NMR spectrometer equipped with a 5 mm z-gradient TCI cryogenic probe and operated at 275 K. Selective excitation of the $H_2O_2$ proton resonance was achieved using a Gaussian 90° pulse of 6 ms duration. The acquisition time for each scan was set to 50 ms, with a recycle delay of 1 ms between successive scans. The identification and quantification of $H_2O_2$ were carried out by comparing the chemical shift positions and signal intensities with those obtained from freshly prepared standard $H_2O_2$ solutions measured under identical conditions. For each sample, 50,000 scans were accumulated to improve the signal-to-noise ratio, enabling a detection limit down to approximately 50 nM.

**EPR Analysis**

Electron paramagnetic resonance (EPR) spectra were obtained using a continuous-wave Bruker ELEXSYS E500 spectrometer, which operates within the X-band frequency range. This apparatus was fitted with a high-sensitivity Bruker ER 4122 SHQ resonator. All measurements were conducted at room temperature under consistent experimental conditions to ensure the reliability and reproducibility of the collected data. Spectra were recorded at a microwave frequency in the X-band region, employing a spectral sweep width of 300 G. Field modulation was applied at a frequency of 100 kHz with a modulation amplitude of 5 G. The microwave power was calibrated to 1.5 mW.

**GC Analysis**

Gas Analyzer (7890A, Agilent Technologies) was equipped with three channels: TCD to separate and detect permanent gases $N_2$, $O_2$, $CO_2$, CO, $H_2$, methane, and ethane, μECD and SCD channels detected other gases. This experiment focuses on the separation of permanent gases through the TCD channel using argon as a carrier gas. The separation of permanent gases was achieved using two simultaneous valves in the system: ten-port gas sampling with back-flush of water to vent, and another six-port column separation of permanent gases into the TCD detector. The TCD channel was tested using a standard gas mixture containing 20% of each $CH_4$ and $CO_2$, 54.5% $N_2$, 5% $H_2$, and 0.5% $O_2$. The headspace on top of the water inside the sample glass bottle capped with a silicone septum was sampled and injected directly into the sampling valve using a 2.5 mL gas-tight syringe (1002 LTN, HAMILTON). The total run time was about 10 min.

**Supporting Information**

The Supporting Information includes video recordings and analytical data showing gas bubble formation and evolution behavior across different electrode materials. **Supporting Movies 1–3** present the experimental setup and real-time visualization of bubble generation in luminol solution under identical conditions using different electrodes. Specifically, **Supporting Movie 1** shows the case of aluminium (Al), **Supporting Movie 2** corresponds to steel, and **Supporting Movie 3** to platinum (Pt). These videos highlight the electrode-dependent chemiluminescence or $HO^\bullet$ radical formation. **Supplementary Figure 1** presents the potentiodynamic polarization test results for different metal surfaces which gives the oxidation rates of these metal surfaces at the solid–water interfaces. In addition, detailed **Supplementary Table 1** contains various experimental observations of chemiluminescence events reported with and without bubbles on different electrode materials.

**Acknowledgments**

The authors are grateful to KAUST Scientific Illustrator Dr. Thom Leach for his contribution to create Figs 1&6 and the graphical abstract. MAE and HM thanks Dr. Krishna Katuri and Prof. Pascal Saikaly (KAUST) for allowing the use of their laboratory's potentiostat (Biologic potentiostat VMP3) for the electrochemical measurements. MAE and HM thank Professor Willem H. Koppenol (ETH Zurich) for productive discussions.

**Competing interests:** The authors declare that they have no competing interests.

**Data and materials availability:** All data needed to evaluate the conclusions in the paper are presented in the paper or the Supplementary Materials.

**Funding:** H.M. acknowledges KAUST for funding (Grant No. BAS/1/1070-01-01).